\documentclass[%
 reprint,
 amsmath,amssymb,
 aps,
]{revtex4-1}

\usepackage{graphicx}
\usepackage{dcolumn}
\usepackage{bm}
\usepackage{bbm}

\usepackage{hyperref}
\hypersetup{%
	pdftitle={Ground and excited states of coupled exciton liquids in electron-hole quadrilayers},%
	pdfauthor={Chao Xu and Michael M. Fogler},%
	pdfpagemode={UseNone},%
	pdfstartview={FitH},%
	breaklinks=true,%
	citecolor=blue,%
	colorlinks=true,%
	linkcolor=blue,%
	urlcolor=blue
}

\renewcommand{\vec}[1]{\mathbf{#1}}

\begin{document}
\title{Ground and excited states of coupled exciton liquids in electron-hole quadrilayers}

\author{Chao Xu}
\affiliation{Physics Department, University of California San Diego, 9500 Gilman Drive, La Jolla, California 92009, USA}

\author{Michael M. Fogler}
\affiliation{Physics Department, University of California San Diego, 9500 Gilman Drive, La Jolla, California 92009, USA}

\date{\today}

\begin{abstract}

Interlayer excitons are bound states of electrons and holes confined in separate two-dimensional layers.
Due to their repulsive dipolar interaction,
interlayer excitons can form a correlated liquid.
If another electron-hole bilayer is present,
excitons from different bilayers can exhibit mutual attraction.
We study such a quadrilayer system by a hypernetted chain formalism.
We compute ground state energies, pair correlation functions, and collective mode velocities as functions of the exciton densities.
We estimate the critical density for the transition to a paired biexciton phase.
For a strongly unbalanced (unequal density) system, the excitons in the more dilute bilayer behave as polarons.
We compute energies and effective masses of such exciton-polarons.

\end{abstract}

\maketitle

\section{Introduction}
\label{sec:introduction}

Indirect exciton or equivalently, interlayer exciton is a neutral quasi-particle in 
a semiconductor nanostructure that contains two parallel layers: one, filled with electrons and the other, with holes.
An interlayer exciton can be created by a photoexcitation of an electron-hole (e-h) pair followed by
separation of the two particles via interlayer tunneling
induced by a strong out-of-plane electric field.
Low-disorder $\rm{GaAs}$-based nanostructures have proved to be particularly suitable
for realization of interlayer exciton systems with tunable density,
long lifetime~\cite{lozovik1976new, butov2001stimulated}, high mobility~\cite{dorow2018high}, and long diffusion length~\cite{hagn1995electric, gartner2006drift, hammack2009kinetics, leonard2009spin, lazic2010exciton, alloing2012nonlinear, lazic2014scalable, finkelstein2017transition}.
In a theoretical analysis of such systems it has been common to treat excitons as composite bosons with no internal dynamics. (Henceforth, ``exciton'' always means interlayer exciton.)
In this approximation each exciton has a permanent dipole moment proportional to the separation of the electron and hole layers.
The interaction of such oriented dipoles located in the same two-dimensional (2D) plane is strictly repulsive.

\begin{figure}
	\center
	\includegraphics[width=3.0in]{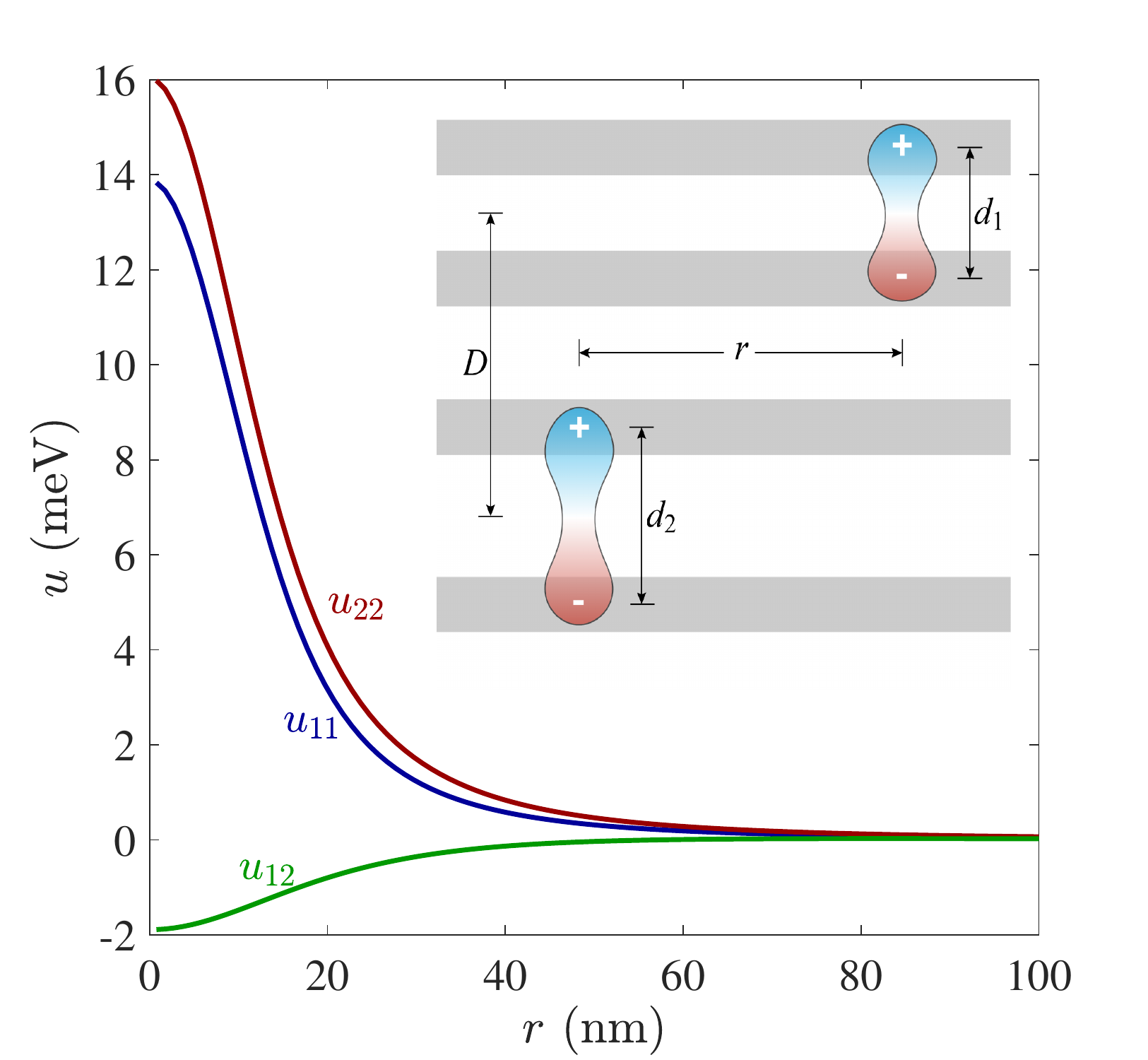}
	\caption{Exciton interaction potentials: $u_{11}$ and $u_{22}$ are intraplane potentials; $u_{12}$ is the interplane potential.
	The inset shows a schematic of the system (see text).
	Parameters: 
	$d_1 = 20\,\mathrm{nm}$, $d_2 = 25\,\mathrm{nm}$, $D = 43\,\mathrm{nm}$, $\kappa = 13$, $c = 5\,\mathrm{nm}$.
	}
	\label{fig:setup}
\end{figure}

Recent experiments~\cite{hubert2019attractive, choksy2021attractive} explored
a more advanced type of $\rm{GaAs}/{\rm{Ga}}{\rm{Al}}{\rm{As}}$ nanostructures containing two e-h bilayers,
as shown schematically in Fig.~\ref{fig:setup}.
In these e-h-e-h quadrilayer systems,
the dipolar interaction between excitons that belong to different e-h bilayers can be of either sign.
This interaction is repulsive at large but attractive at small in-plane distances $r$ between the excitons, see the curve labeled $u_{12}$ in Fig.~\ref{fig:setup}.
Experimental evidence for the interlayer attraction~\cite{hubert2019attractive, choksy2021attractive} was deduced from the dependence of the exciton photoluminescence energy 
and the exciton density distribution~\cite{choksy2021attractive} on
the separately controlled average exciton densities in the two bilayers.
Motivated by these experiments, in this paper we undertake a quantitative analysis of the ground state and excitations of
an e-h-e-h quadrilayer system.

Our study is a continuation of extensive prior theoretical work on e-h bilayers
and 2D dipolar bosons.
For example, the phase diagram of
a single e-h bilayer for the case where electrons and holes have equal densities $n$ and masses $m_e = m_h$
has been explored by
several Monte-Carlo simulations~\cite{de2002excitonic, tan2005exciton, lee2009exciton, maezono2013excitons}.
Such simulations are considered to be the most reliable tool for
the case of strong correlations.
These studies have shown that the excitons are stable when $n$ is below a certain threshold (Mott critical density)
$n_{c2} = c_2 a_e^{-2}$. Here $c_2 \sim 0.02$ is a numerical coefficient that
depends on the dimensionless ratio $d / a_e$ of the e-h separation $d$ and the electron Bohr radius $a_e = \hbar^2 \kappa / m_e e_0^2$ with $\kappa$ and $e_0$ being the dielectric constant and the elementary charge, respectively.
Neglecting internal dynamics of excitons is justified if $n \ll n_{c2}$.
The ground state of the system is determined by the competition between
the kinetic energy of the excitons and their dipole repulsion that scales as $d^2$.
In the experimentally relevant regime~\cite{hubert2019attractive, choksy2021attractive}
$d / a_e \sim 0.3$, excitons are expected to form a strongly correlated Bose liquid (which is a superfluid).
At much larger or much smaller $d$'s, other phases of exciton matter, e.g., exciton solid are possible.

Phase diagram of 2D dipolar bosons inferred from Monte-Carlo simulations shows
a close correspondence to that of the e-h bilayer in regards to the position of
the liquid-solid phase boundary~\cite{astrakharchik2007quantum, Astrakharchik2007gsp, bucher2007strongly}.
Numerical results for the ground-state energy and density correlation function of 2D dipolar bosons have been conveniently summarized in~\cite{astrakharchik2009equation} and we will use some of them in this work.
The excitation spectra of such systems~\cite{astrakharchik2007quantum, hufnagl2011roton, abedinpour2012theory} have also been studied.
Additional work in this subject area includes investigations of the thermal melting of dipolar solids~\cite{Mora2007melting} and the scattering-length instability~\cite{bortolotti2006scattering}.
The latter has implications for the normal-superfluid transition. 
For systems of dipoles whose orientation is tilted away from the vertical, a unidirectional density wave (stripe) phase was predicted to appear~\cite{macia2012excitations, macia2014phase}. 

\begin{figure}
	\center
	\includegraphics[width=3.0in]{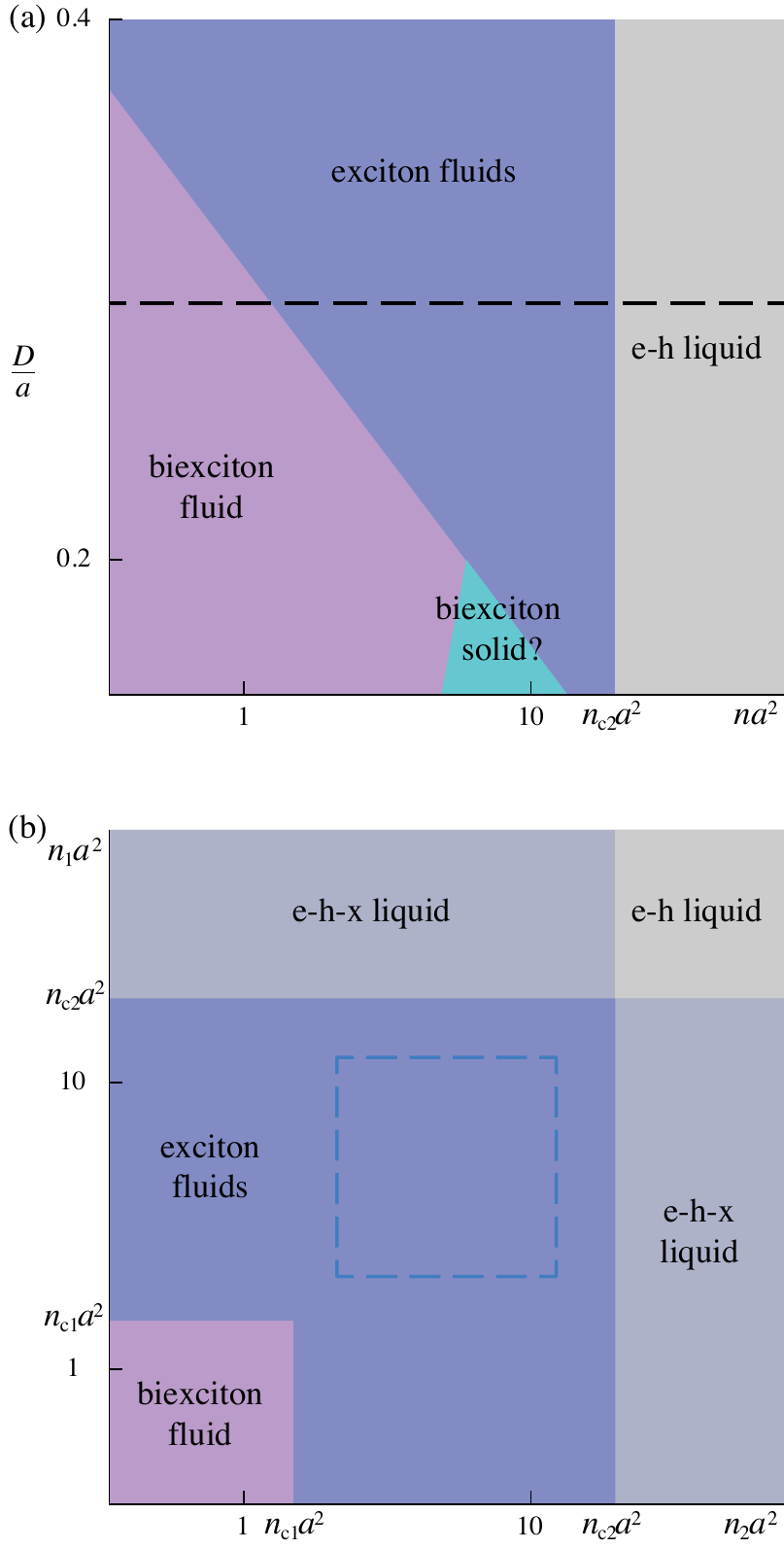}
	\caption{(a) Schematic zero-temperature phase diagram of a symmetric two-plane exciton system,
	 $n_1 = n_2 \equiv n$. (b) Schematic phase diagram for arbitrary exciton densities $n_1$, $n_2$. The interplane distance $D$ is fixed at the position of the dashed line in panel (a); ``e'', ``h'', and ``x'' stand for electron, hole, and exciton,
	respectively.
	The biexciton fluid contains some unpaired excitons if $n_1 \neq n_2$.
	The dashed square indicates the $n_1$--$n_2$ range plotted in Fig.~\ref{fig:energy}.
}
	\label{fig:phasediagram}
\end{figure}

The most directly relevant to our work are Monte-Carlo simulations done
for bilayer systems of magnetic dipolar bosons~\cite{macia2014single, filinov2016correlation, cinti2017phases},
which have essentially the same interaction law as excitons.
Based on these studies, we can surmise the following structure of the zero-temperature phase diagrams of
e-h-e-h quadrilayers.
To avoid confusion we will use the term ``plane'' in relation to excitons
and ``layer'' for electrons or holes.
(Hence, a plane is made of a pair of adjacent layers.)
Let us start with a symmetric case, i.e., two parallel planes
each filled with excitons of dipole moment $e_0 d$ and density $n$.
The planes are separated by a distance $D > d$.
The dimensionless parameters of the problem
are obtained by multiplying $D$, $n$, and $d$ by appropriate powers of
the dipole length $a = d^2 / a_x$, where $a_x = \hbar^2 \kappa / m e_0^2 = a_e m_e / m$ is the effective exciton Bohr radius and $m = m_e + m_h$ is the exciton mass.
As shown schematically in Fig.~\ref{fig:phasediagram}(a),
if $D / a$ and $n a^2$ are small, so that the mean in-plane exciton distance $n^{-1 / 2}$ is relatively large, the interplane attraction of excitons favors paired phases.
This means that excitons from the opposite planes bind into biexcitons
if $n$ is less than some critical density $n_{c1} = n_{c1}(D)$.
The biexcitons would typically form a correlated liquid but
a small region of the solid phase is also possible. 
When $D / a$ or $n a^2$ is large, the intraplane repulsion dominates over the interplane attraction,
so that the unpaired exciton fluids are stable.
However, the excitons should dissociate
once $n$ increases beyond the Mott critical density, which is
the rightmost part of the phase diagram.
The rectilinear phase boundaries in Fig.~\ref{fig:phasediagram}(a) are meant
to be schematic only.

In principle, more exotic phases of matter are possible in this system.
For example, intraplane exchange interaction and spin-related effects
can become important at high exciton density.
In this paper, we focus on the moderate
and low-density regimes, and so we ignore such effects.
Note that the interplane exchange interaction can
be usually neglected at all exciton densities.

The dashed line in Fig.~\ref{fig:phasediagram}(a) indicates the interplane distance representative of the experiments cited above~\cite{hubert2019attractive, choksy2021attractive}.
In such experiments, the exciton density $n$ can be controlled by photoexcitation power; however,
very small or very larger $n$ are usually difficult to access due to, respectively, disorder and
heating effects.
Therefore, the biexciton and the unpaired exciton fluids are the most relevant phases.

In practice, excitons in the two planes may have different densities $n_1$ and $n_2$.
In Fig.~\ref{fig:phasediagram}(b) we present a crude phase diagram for a representative range of $n_1$ and $n_2$.
This diagram is based on the notion that plane $k$ contains unbound e-h pairs if $n_k > n_{c2}$ and unbound excitons if $n_k > n_{c1}$.
In this work we estimate $n_{c1}$ and compute other basic many-body properties
of the system as functions of $n_1$ and $n_2$.
To do so we use the hypernetted chain (HNC) method,
which is only slightly less accurate than the Monte-Carlo simulations.

The remainder of this article is organized as follows. In Sec.~\ref{sec:setup}, we define the model we study.
In Sec.~\ref{sec:ground-state}, we report the ground-state energies and
sound velocities calculated for relatively high exciton densities.
We verify the accuracy of our HNC method by comparing it with the available Monte-Carlo simulations for the single-plane problem.
In Sec.~\ref{sec:low-density}, we qualitatively discuss the low-density paired phase.
In Sec.~\ref{sec:impurity}, we study the regime where only one of the planes is dilute.
Here the analysis can be made more quantitative
using the analogy to the polaron problem.
We compute the energies and effective masses of such exciton-polarons.
We give concluding remarks in Sec.~\ref{sec:conclusions}.
Additional calculations and technical details are presented in the Appendix.

\section{Model}
\label{sec:setup}

Our model is sketched in the inset of Fig.~\ref{fig:setup}.
The white strips indicate the tunneling barrier regions, which are classically forbidden for the carriers. 
The dark strips represent the layers populated alternatively by electrons ($-$) and holes ($+$).
We assume that all the electrons and holes are paired into excitons
and that the pairing occurs only in the two top and two bottom layers
that have center-to-center separations $d_1$ and $d_2$, respectively.
We ignore the possibility of exciton formation by binding carriers from the two middle layers.
This should be legitimate if the corresponding interlayer distance $D - (d_1 + d_2) / 2$ is sufficiently large.
We also ignore any internal dynamics of excitons,
which allows us to treat the quadrilayer as two coupled planes of excitons.
Here, as in Sec.~\ref{sec:introduction}, the term ``layer'' pertains to electrons and holes, while
``plane'' describes a 2D sheet of excitons.
Specifically, the exciton plane $\alpha = 1$ ($2$) represents the two top (two bottom) layers seen as a unit.
The effective Hamiltonian we study is the sum of the kinetic and
interaction energies:
\begin{align}
	H &= H_1 + H_2 + H_{12}\,,
	\label{eqn:H}\\
	H_{\alpha} &= -\frac{\hbar^2}{2m} \sum_{i = 1}^{N_{\alpha}} \nabla^2_i
	+ \sum_{i < j} u_{\alpha \alpha}\left(\vec{r}^i_\alpha - \vec{r}^j_\alpha\right),
	\\
	H_{12} &= \sum_{i = 1}^{N_1}
	\sum_{j = 1}^{N_2} u_{12}\left(\vec{r}^i_1 - \vec{r}^j_2\right).
\end{align}
Here $\vec{r}^i_\alpha$'s are the exciton coordinates, $N_\alpha$ is their number,
and $n_\alpha = N_\alpha / \Omega$ is their density in plane $\alpha$; $\Omega$ is the system area.
We assume that the effective mass $m$ of the excitons is the same in the two planes
(approximately $0.20$ of the bare electron mass in $\rm{GaAs}/{\rm{Ga}}{\rm{Al}}{\rm{As}}$ nanostructures~\cite{choksy2021attractive}). 

In order to model the effective interaction potentials $u_{\alpha \beta}(r)$ we need to know
the charge distribution of the exciton.
In principle, it can be determined by solving the two-body e-h binding problem numerically.
However, the result depends on many microscopic details, such as
the thickness of the layers, the electric field, and the properties of the tunneling barriers~\cite{Szymanska2003ebi, Sivalertporn2012dai}.
As a simpler alternative, we assume that the charge density distribution of every electron (hole) in
a given layer $j$ is an isotropic Gaussian:
\begin{equation}
	|\psi(r, z)|^2 \propto \exp\left[-\frac{(z - z_j)^2 + r^2}{2 c^2}\right]\,.
\label{eqn:Gaussian}	
\end{equation}
Here $r$ is the in-plane distance of the particle from the center of the exciton and
$z$ is its out-of-plane coordinate, with $z_j$ being the midpoint $z$-coordinate of layer $j$. 
The Coulomb interaction energy of two such Gaussians of charge $e_0$ each has the following analytic form
\begin{equation}
	v(r, z) = \frac{e_0^2}{\kappa \sqrt{r^2 + z^2}}\,
	 {\rm{erf}}\left(\frac{\sqrt{r^2 + z^2}}{2 c}\right),
\label{eqn:v}
\end{equation}
which is familiar from the Ewald summation method.
The second factor in Eq.~\eqref{eqn:v},
containing the error function $\mathrm{erf}(x)$, smooths out the short-range divergence of the
Coulomb potential.
Later we will need the 2D Fourier transform of $v(r, z)$ with respect to $r$,
which is given by
\begin{align}
\tilde{v}(k, z) &\equiv \int v(r, z) e^{-i \vec{k}\cdot \vec{r}} d^2 r
\label{eqn:v_q_def}\\
       &= \frac{e_0^2}{\kappa}\, \frac{\pi}{k} \sum_{\sigma = \pm 1}
e^{k^2 c^2 + \sigma k z}
	{\rm{erfc}}\left(k c + \frac{\sigma z}{2 c}\right),
	\label{eqn:v_q}
\end{align}
where $k = |\vec{k}|$ and $\mathrm{erfc}(x) = 1 - \mathrm{erf}(x)$.

In our approximate model the charge density distribution of the exciton consists of two oppositely-charged Gaussians [Eq.~\eqref{eqn:Gaussian}] aligned in-plane.
Accordingly,
the interplane $u_{12}(r)$ and intraplane $u_{\alpha \alpha}(r)$ exciton interaction potentials are given by
\begin{align}
    u_{12}(r) &= v\left(r, D + \frac{d_1 - d_2}{2}\right)
               + v\left(r, D - \frac{d_1 - d_2}{2}\right)
\notag\\
              &- v\left(r, D + \frac{d_1 + d_2}{2}\right)
               - v\left(r, D - \frac{d_1 + d_2}{2}\right),
\label{eqn:u_12}\\
	u_{\alpha \alpha}(r) &= 2 v\left(r, 0\right) - 2 v\left(r, d_\alpha\right).
\label{eqn:u_k}
\end{align}
These potentials are plotted in Fig.~\ref{fig:setup} for representative parameter values~\cite{choksy2021attractive}. 
In this example, $u_{11}(r)$ and $u_{22}(r)$ are always positive while $u_{12}(r)$ is negative at $r < 57\,\mathrm{nm}$,
see also Fig.~\ref{fig:HNCvsPolaron}(d) below.
All the potentials decay as $u_{\alpha \beta}(r) \propto r^{-3}$ at large $r$, which classifies them as short-range interactions.
The width $c$ of the Gaussian in Eq.~\eqref{eqn:Gaussian} affects mainly the short-distance behavior of the intraplane potentials $u_{\alpha \alpha}(r)$.
Since the excitons in the same plane strongly avoid each other (Sec.~\ref{sec:ground-state}),
this adjustable parameter has only a minor influence on the many-body properties.
The properties of our main interest are the ground-state energy and the low-energy excitation spectrum. In the following Section we discuss methods we employ to study them.

\section{HNC formalism}
\label{sec:HNC}

The primary many-body quantities we were able to compute include the pair correlation functions (PCFs) $g_{\alpha \beta}(r)$,
the structure factors $S_{\alpha \beta}(k)$,
and the energy density per unit area $e = e(n_1, n_2)$.
The PCF is defined by
\begin{equation}
	g_{\alpha \beta}(r) = \frac{\Omega}{N_\alpha N_\beta} \sum_{i = 1}^{N_\alpha} \sum_{j = 1}^{N_\beta} \left\langle \delta\left(\vec{r}^i_\alpha - \vec{r}^j_\beta - \vec{r}\right)\right\rangle
	- \frac{\delta_{\alpha \beta}}{n_\alpha} \delta\left(\vec{r}\right).
	\label{eqn:g_lm}
\end{equation}
The structure factor is $S_{\alpha \beta}(k) = \delta_{\alpha \beta} + \sqrt{n_\alpha n_\beta}\, \tilde{h}_{\alpha \beta}(k)$ where $h_{\alpha \beta}(r) \equiv g_{\alpha \beta}(r) - 1$.
The tilde in $\tilde{h}_{\alpha \beta}$ denotes the 2D Fourier transform,
as in Eq.~\eqref{eqn:v_q_def}.

The energy density can be expressed in terms of the PCF.
The simplest result is obtained within the mean-field approximation, $g_{\alpha \beta} = 1$, $S_{\alpha \beta} = \delta_{\alpha \beta}$ that neglects correlations.
This mean-field energy density is
\begin{equation}
	e_\mathrm{mf}(n_1, n_2) = \frac12 \sum_{\alpha \beta} \tilde{u}_{\alpha \beta}(0) n_\alpha n_\beta\,.
	\label{eqn:e_mean_field}
\end{equation}
For our model interaction potentials, we find
\begin{align}
	\tilde{u}_{\alpha \alpha}(0) &= \int u_{\alpha \alpha}(r) d^2 r
	= 2 \lim_{q \to 0 } [\tilde{v}(q, 0) - \tilde{v}(q, d_\alpha)]
	\label{eqn:u_k_integrated_def}\\
	&= \frac{4\pi e_0^2}{\kappa}\left[d_\alpha\, \mathrm{erf}\left(\frac{d_\alpha}{2 c}\right)
	- \frac{2 c}{\sqrt{\pi}} \left(1 - e^{-\frac{d_\alpha^2}{4 c^2}} \right) \right],
	\label{eqn:u_k_integrated}
\end{align}
so that in the zero-thickness limit, $c \to 0$, we get
\begin{equation}
\tilde{u}_{\alpha \alpha}(0) \to \frac{4\pi e_0^2}{\kappa}\, d_\alpha\,,
\qquad
\tilde{u}_{12}(0) \to 0\,.
\label{eqn:u_k_capacitor}
\end{equation}
This simplified expression yields the ``capacitor formula''
\begin{equation}
	e_\mathrm{cap}(n_1, n_2) = \frac{2\pi e_0^2}{\kappa}\, \left(n_1^2 d_1 + 
	n_2^2 d_2\right),
	\label{eqn:e_mean_field_two_planes}
\end{equation}
which is so named because it resembles the total energy of two parallel-plate capacitors of dielectric thickness $d_1$ and $d_2$.
There is no $n_1 n_2$ term in this quadratic form since
such parallel-plate capacitors do not produce external electric field, and so do not interact.
Equation~\eqref{eqn:e_mean_field_two_planes} implies that any appreciable effect of interplane interactions can arise only from correlations.

\begin{figure}
	\center
	\includegraphics[width=3.0in]{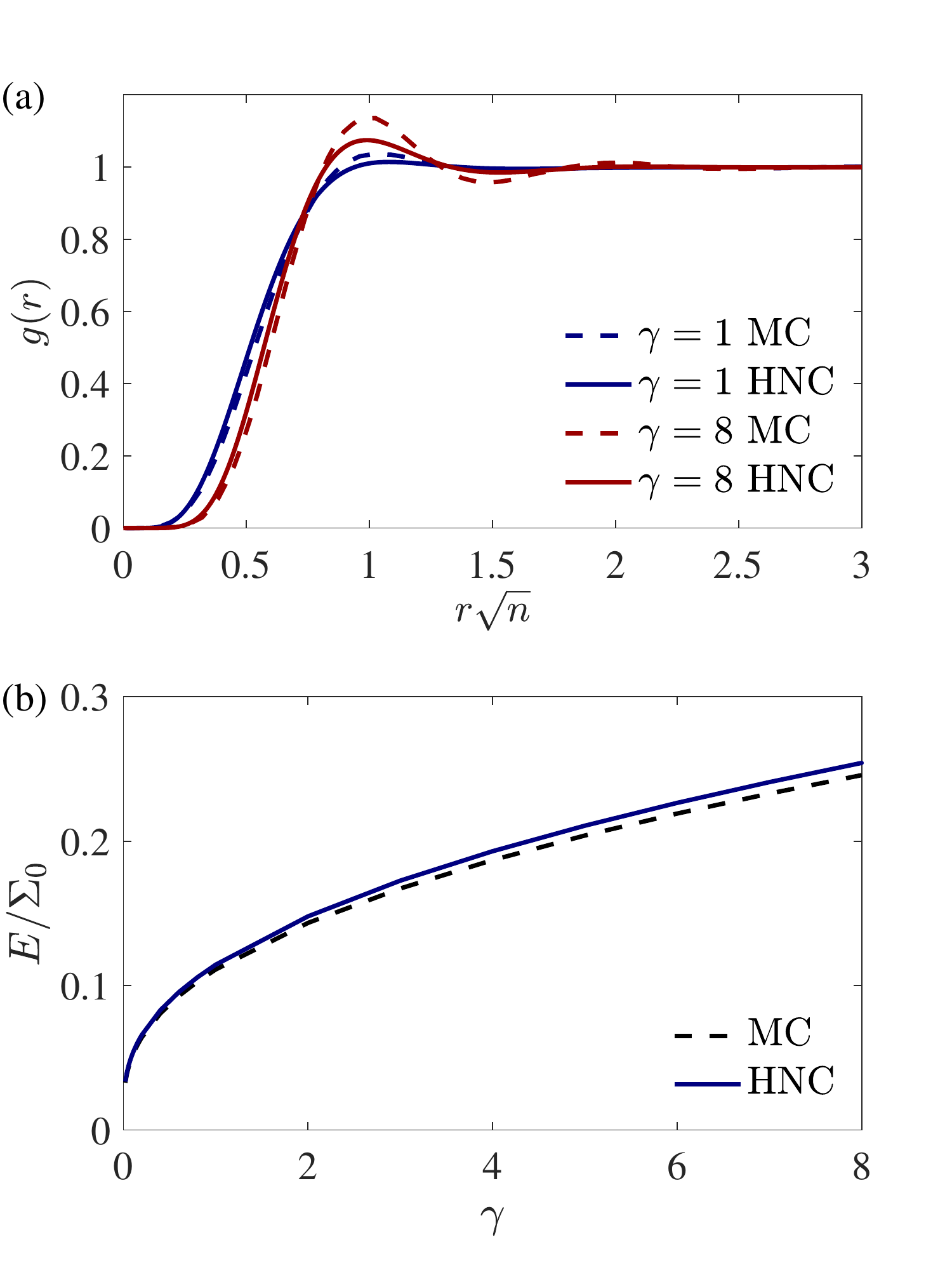}
	\caption{Comparison of our HNC calculations with prior Monte-Carlo results~\cite{astrakharchik2007quantum, Astrakharchik2007gsp, bucher2007strongly} for a model system of single-plane dipolar bosons.
		(a) PCF computed for two values of $\gamma = n a^2$ representative of
		correlated exciton liquids in ${\rm{GaAs}}$~\cite{choksy2021attractive}.
		Note that the liquid-solid transition occurs at $\gamma \approx 300$.		
		(b) Energy per particle $E$ in units of the capacitor self-energy $\Sigma_0$ (see text) as a function of $\gamma$ for fixed $d = 6 a_x$.}
	\label{fig:HNCvsMC}
\end{figure}

To go beyond the mean-field theory we employed the zero-temperature HNC (more precisely, HNC/$0$) method
for multi-species systems~\cite{chakraborty1982structure, Chakraborty1982VTO}.
In this method, the ground-state wavefunction $\Psi$ is assumed to be in the
Jastrow-Feenberg product form:
\begin{equation}
    \Psi = \prod_{\alpha = 1, 2} \, \prod_{i < j}
     f_\alpha\left(\vec{r}^i_\alpha - \vec{r}^j_\alpha\right)
 \prod_{i,\, j} f_{12}\left(\vec{r}^i_1 - \vec{r}^j_2\right).
\label{eqn:Psi}
\end{equation}
Functions $f_1$, $f_2$, $f_{12}$ obey a set of nonlinear equations
\begin{equation}
	\left( -\frac{\hbar^2}{m} \nabla^2 + u_{\alpha\beta}(r) + w_{\alpha\beta}(r)\right ) \sqrt{g_{\alpha\beta}(r)} = 0\,,
\label{eqn:g_equation}
\end{equation}
where the so-called induced potentials $w_{\alpha\beta}(r)$ are defined
via their Fourier transforms
\begin{equation}
	\tilde{w}_{\alpha\beta}(k) = \frac{\epsilon(k)}{2 \sqrt{n_{\alpha}n_{\beta}}}
	\left[3 \mathrm{I} - 2 \mathrm{S} - (\mathrm{S}^{-1})^2 \right]_{\alpha\beta},
\label{eqn:w}
\end{equation}
$\mathrm{S}$ is the $2 \times 2$ matrix made of $S_{\alpha \beta}(k)$,
$\mathrm{I}$ is the identity matrix, and
\begin{equation}
	\epsilon(k) = \frac{\hbar^2 k^2}{2 m}
\label{eqn:epsilon}
\end{equation}
is the bare single-particle energy.
These equations can be solved numerically by iterations~\cite{chakraborty1982structure, Chakraborty1982VTO}.
The energy density of the system is then calculated from~\cite{Chakraborty1982VTO}
	\begin{align}
		e &= \frac{1}{2}\int d^2 r \sum_{\alpha \beta} n_\alpha n_\beta \left[
		g_{\alpha \beta} u_{\alpha \beta}
		+  \frac{\hbar^2}{m} (\nabla \sqrt{g_{\alpha \beta}})^2\right]\\
		&- \frac{1}{4} \int\frac{d^2k}{(2\pi)^2} \epsilon(k)
\,\mathrm{tr} \left(3 \mathrm{I} - 3 \mathrm{S} + \mathrm{S}^2 - \mathrm{S}^{-1}\right).
	\end{align}
The performance of the HNC method has been previously shown to be very good~\cite{abedinpour2012theory} in a single-plane system, $N_1 = 0$.
The corresponding ground-state wavefunction is obtained
from Eq.~\eqref{eqn:Psi} by dropping $f_1$ and $f_{12}$ terms while
in the HNC equations one needs to set $S_{11} = 1$, $S_{12} = 0$, and solve for $g_{22}$ only.
To simplify notations, we also drop the subscripts in $n_2$, $S_{22}$, $g_{22}$, \textit{etc}.
In this calculation the point-dipole limit $u(r) = e_0^2 d^2 / r^3$
was used for which Monte-Carlo results are available~\cite{astrakharchik2007quantum, Astrakharchik2007gsp, bucher2007strongly}.
Such a system can be characterized by the dimensionless interaction strength
$\gamma = n a^2$, where $a = d^2 / a_x$ is the dipole length introduced in Sec.~\ref{sec:introduction}.
Note that $\gamma \sim 4$ for density $n \sim 1.0 \times 10^{10}\, \mathrm{cm}^{-2}$ and dipole length $a \sim 200\, \mathrm{nm}$
typical of ${\rm{GaAs}}$ devices. 
We repeated this benchmark calculation on a denser grid of density points
in the practical range $\gamma < 8$ of moderately strong interactions.
In Fig.~\ref{fig:HNCvsMC} we present the results for the PCF
and the energy per particle $E = e(n) / n$ in units of 
the ``capacitor self-energy'' $\Sigma_0 = ({4\pi e_0^2}/{\kappa}) ({n d})$.
From Fig.~\ref{fig:HNCvsMC}, we can see that the HNC and the more accurate Monte-Carlo methods are indeed in a good agreement.
Both methods predict that the particles strongly avoid each other at distances shorter than the average intraplane spacing $n^{-1 / 2}$,
which enables them to reduce the system energy significantly below the mean-field value.

\section{High to moderate exciton densities}
\label{sec:ground-state}

\begin{figure}
	\center
	\includegraphics[width=3.5in]{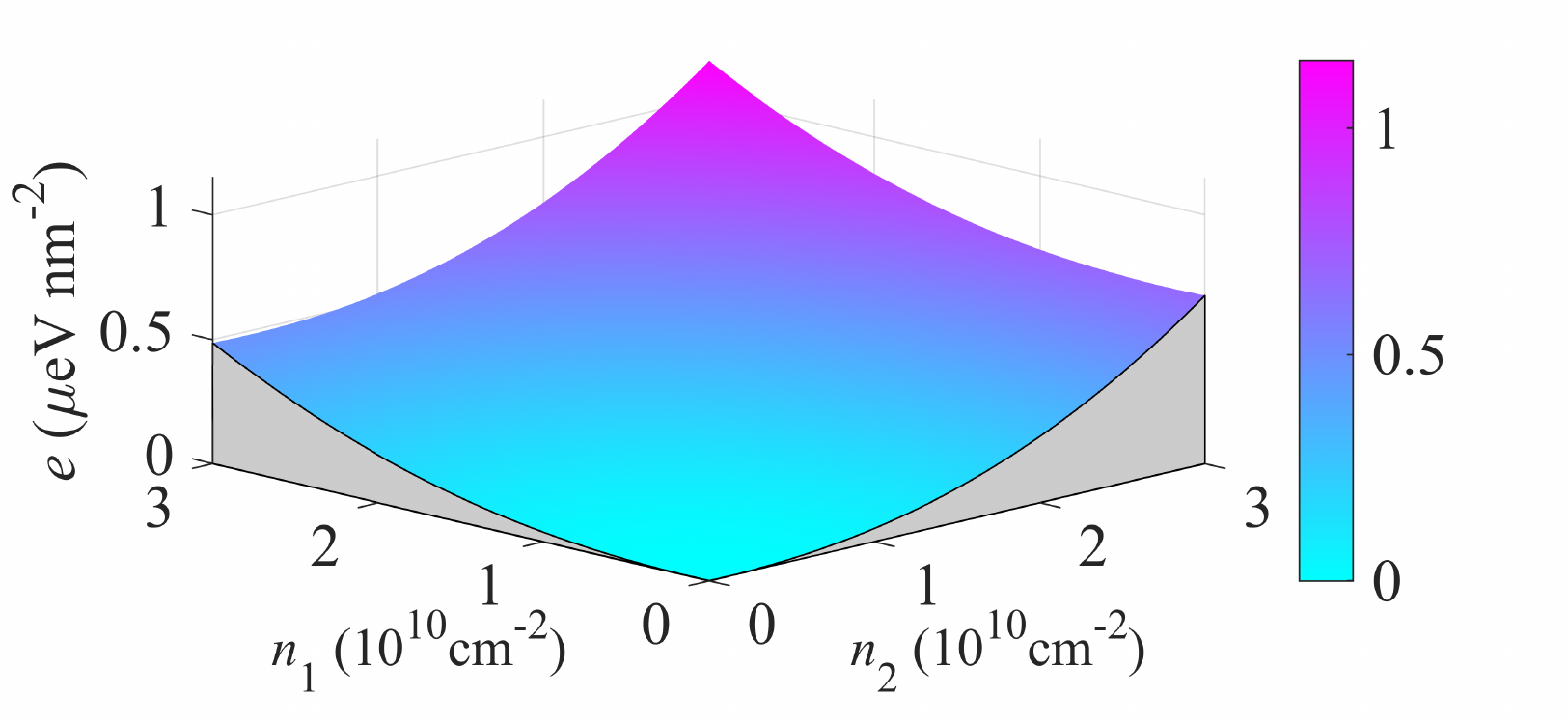}
	\caption{Energy density given by the HNC.
	}
	\label{fig:energy}
\end{figure}

Let us turn to our main subject, the two-plane exciton system.
Including all three correlation factors $f_1$, $f_2$, $f_{12}$,
and solving the HNC equations for a range of densities
$n_1, n_2$, we arrived at the results
presented in Fig.~\ref{fig:energy}.
The geometrical and physical parameters used in the calculations
are specified in the caption of Fig.~\ref{fig:setup}.
Qualitatively, the behavior of the obtained energy density $e(n_1, n_2)$ resembles the predictions of the mean-field theory.
However, the energy density is 
greatly reduced compared to
Eq.~\eqref{eqn:e_mean_field_two_planes}
and this reduction is stronger at small $n_1$, $n_2$,
as in the single-plane test case, Fig.~\ref{fig:HNCvsMC}(b).
The effect of many-body correlations can be seen more clearly in derivatives of function $e(n_1, n_2)$,
which are also more directly connected to quantities measured in experiments.
For example, the first derivatives, i.e.,
the chemical potentials
\begin{equation}
\Sigma_\alpha = \frac{\partial e}{\partial n_\alpha},
\qquad \alpha = 1, 2,
\label{eqn:Sigma_l}	
\end{equation}
are related to the exciton photoluminescence energies.
Calculations of $\Sigma_\alpha$ and their comparison with experiments
in ${\rm{GaAs}}$ systems have been reported in Ref.~\cite{choksy2021attractive}.
Here we focus on the second derivatives $\partial^2 e / \partial n_\alpha \partial n_\beta$ of the energy density,
which determine another physical observable: the spectrum of low-energy excitations.

Recall that the elementary excitations of a single-component Bose liquid with short-range interactions are phonons with a linear dispersion at small momentum:
$E(k) \simeq \hbar v k$.
Our two-plane system supports two phonon modes
that represent coupled oscillations of the exciton densities.
Their velocities $v_j$, where $j = 1, 2$, satisfy the condition that
$m v_j^2$ are the eigenvalues of a $2 \times 2$ matrix with
elements 
\begin{equation}
	\sqrt{n_\alpha n_\beta}\, \frac{\partial^2 e}{\partial n_\alpha \partial n_\beta}\,.
\label{eqn:v_from_e}	
\end{equation}
We define $v_2$ ($v_1$) to be the larger (smaller) of the two velocities.

\begin{figure}
	\center
	\includegraphics[width=3.5in]{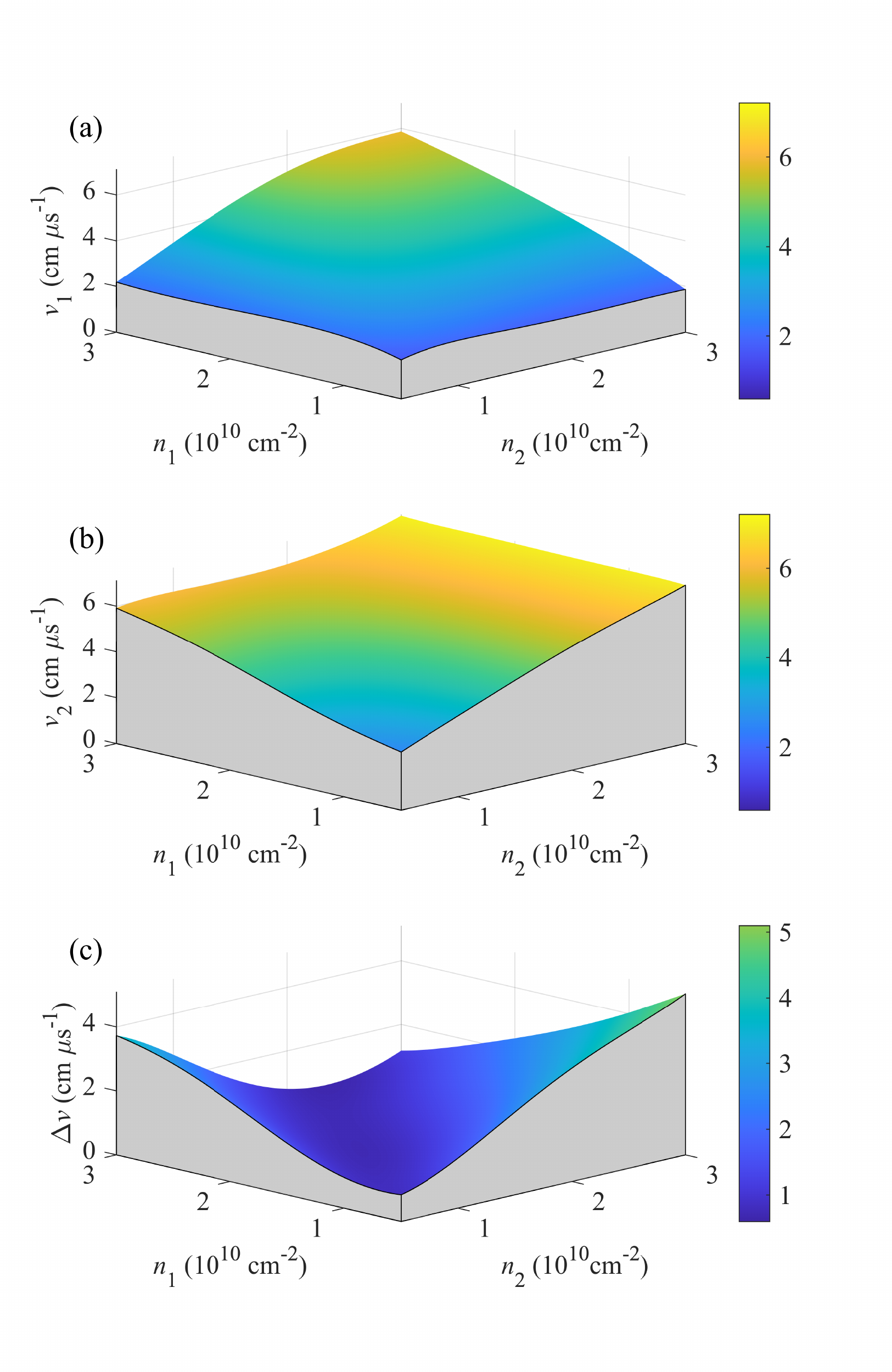}
	\caption{Sound velocities as functions of the exciton densities.
		(a) $v_1$, the slower velocity (b) $v_2$, the faster velocity (c) $\Delta v = v_2 - v_1$, the difference of the two.}
	\label{fig:spec}
\end{figure}

Within the mean-field theory, the second derivatives in question are equal to the plane-integrated interaction potentials, e.g.,
\begin{equation}
	\frac{\partial^2 e_\mathrm{mf}}{\partial n_1 \partial n_2} = \tilde{u}_{1 2}(0)\,.
	\label{eqn:e2n2_mean-field}	
\end{equation}
Since $\tilde{u}_{1 2}(0)$ is small, the mean-field theory predicts that the two phonon modes are
nearly decoupled.
The HNC method should give a superior approximation for the energy density,
and thus for the coupling (``mode repulsion'') of the sound velocities.
In Fig.~\ref{fig:spec}(a, b) we plotted
$v_1$ and $v_2$ deduced from the HNC for the same parameters as in
Figs.~\ref{fig:setup} and \ref{fig:energy}.
The typical magnitude of the sound velocities is a few times $10^6\,\mathrm{cm / s}$.
For reference, the kinetic energy of a free exciton 
with such a velocity is of the order of several degrees $\mathrm{K}$.
Figure~\ref{fig:spec}(b) shows that as $n_1$ increases at fixed $n_2$, the 
dependence of $v_2$ on $n_1$ flattens out.
Similar trend is observed when $n_2$ increases at fixed $n_1$.
This occurs because the interplane correlations weaken at large densities.
To reveal the mode coupling we plotted the difference $\Delta v = v_2 - v_1$ in
Fig.~\ref{fig:spec}(c).
The mode repulsion is evidenced by the appreciable value of $\Delta v$ at the bottom
of the deep trough running roughly diagonally through this plot.

The full momentum dependence of particle density excitation spectra $E_j(k)$ can reveal further
information about correlations.
The mean-field (Gross-Pitaevskii) theory predicts
\begin{equation}
\begin{split}
	E_\mathrm{mf}^2(k) &= \epsilon^2(k) + \epsilon(k) (n_1 \tilde{u}_{11} + n_2 \tilde{u}_{22})
		\\
	& \pm \epsilon(k) \sqrt{(n_1 \tilde{u}_{11} - n_2 \tilde{u}_{22})^2 + 4 n_1 n_2 \tilde{u}^2_{12}}
	\,.
\end{split}
\label{eqn:E_mf}
\end{equation}
On the other hand, the exact $E_j(k)$ are determined by the poles of the dynamic structure factor
(more generaly, by the regions of the $k$--$E$ space where this factor has a nonzero imaginary part).
Unfortunately, the dynamical structure factor is not available from the HNC.
Following previous work~\cite{Astrakharchik2007gsp, hufnagl2011roton, macia2012excitations, abedinpour2012theory}
we estimate $E_j(k)$ from the static structure factors
$S_{\alpha \beta}(k)$ using the Bijl-Feynman approximation (BFA)~\cite{mahan2013many}.

\begin{figure}
	\center
	\includegraphics[width=3.5in]{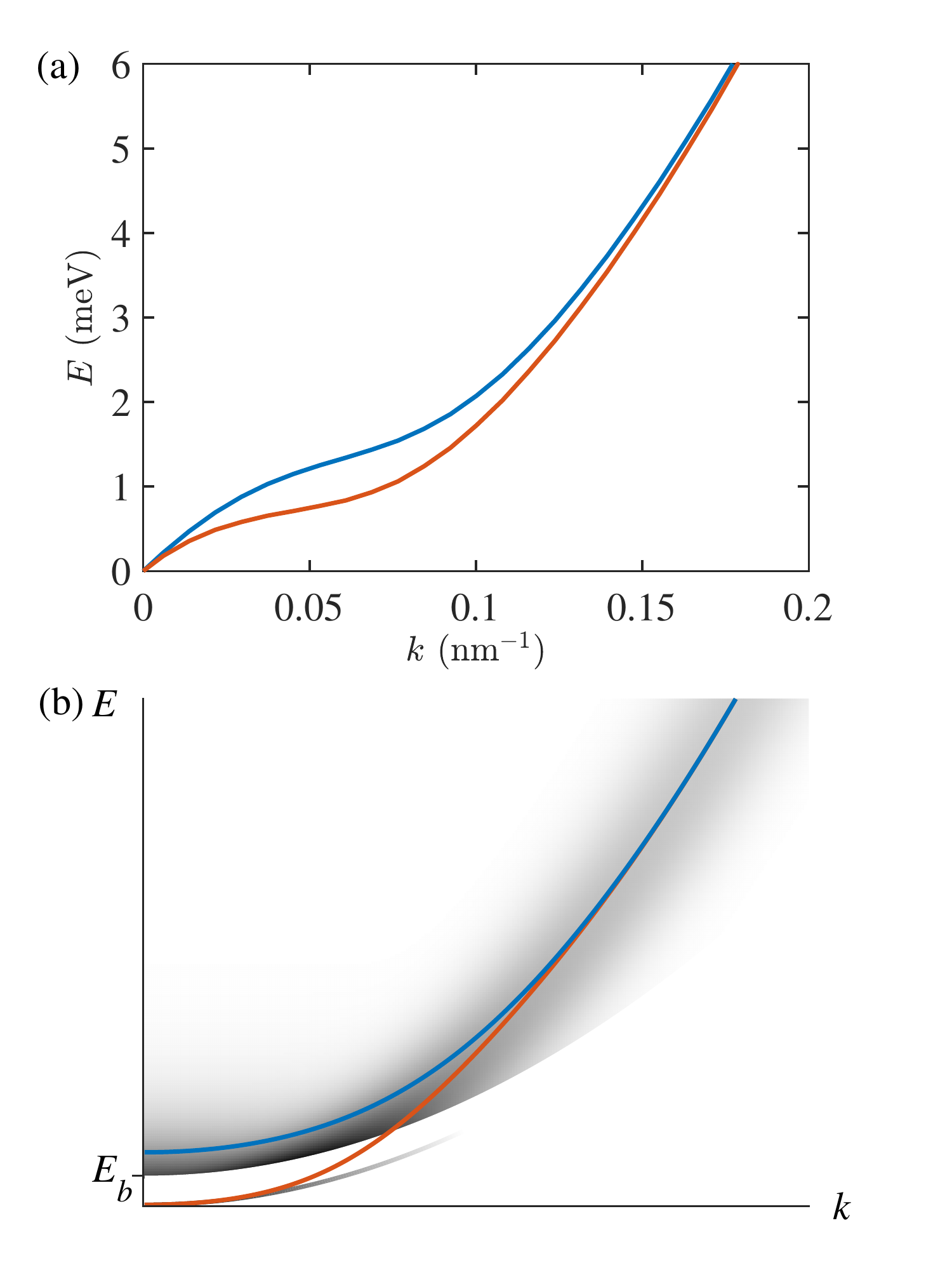}
	\caption{Collective mode spectra.
		(a) Mode dispersions for $n_1 = n_2 = 2.0\times 10^{10}\, \mathrm{cm}^{-2}$ computed using the BFA.
		(b) A simplified sketch of the excitation spectrum in the biexciton phase. The grayscale shading represents the true spectral weight and the solid lines indicate the BFA.}
	\label{fig:dilute_spec}
\end{figure}

The BFA can be derived by diagonalizing the Hamiltonian in a subspace of density-wave states
\begin{equation}
|\vec{k}_\alpha \rangle \equiv \frac{1}{\sqrt{N_\alpha^{\phantom{\dagger}}}}
|\rho^\dagger_\alpha(\vec{k}) | 0 \rangle,
\label{eqn:density_wave}
\end{equation}
where $\rho_\alpha(\vec{k}) = \sum_{j = 1}^{N_\alpha} e^{-i \vec{k}\cdot \vec{r}_\alpha^j}$ is the
density operator in plane $\alpha$.
The key to the derivation are the following identities~\cite{mahan2013many}
\begin{gather}
\langle \vec{k}_\alpha | \vec{k}_\beta \rangle = S_{\alpha \beta}(k) + \sqrt{N_\alpha N_\beta}\, \delta_{\vec{k}= 0},
\label{eqn:rho_rho}\\
\langle \vec{k}_\alpha | H' |\vec{k}_\beta \rangle = \delta_{\alpha \beta} S_{\alpha \alpha}(k)
	\epsilon(k),
\label{eqn:rho_H_rho}
\end{gather}
where $H' = H - e \Omega$ is the Hamiltonian with the ground-state energy subtracted.
Using these relations, one can obtain and easily solve a $2 \times 2$ matrix eigenvalue problem.
The result is
\begin{equation}
	E_{1, 2}(k) = \frac{2 \epsilon(k)}{S_{11} + S_{22}
		\mp \sqrt{(S_{11} - S_{22})^2 + 4 S_{12}^2}}
\label{eqn:E}
\end{equation}
and a representative BFA spectrum is shown in Fig.~\ref{fig:dilute_spec}(a).
This Figure demonstrates that the dispersion of the two modes remains accurately linear up to $k \sim 1 / \sqrt{n_j}$.
At larger $k$ the deviations from the linearity start to be noticeable.
The dispersions subsequently develop plateau-like structures
indicative of strong short-range correlations in the system.
At still larger $k$, the two modes merge together,
as they both approach the free-particle limit
$E_j(k) \to \epsilon(k)$.
Note that the BFA is somewhat misleading 
because the exact excitation spectrum at finite $k$ is not
confined to two dispersion lines of infinitesimal width.
It is known that the excitations span instead a continuum of energies and the BFA shows only
the center-of-gravity (the first moment) of this continuum.

\section{Low-density paired phase}
\label{sec:low-density}

At low densities our HNC simulations were hindered by the lack of convergence.
We attribute this to the proximity of the paired superfluid phase,
see Sec.~\ref{sec:introduction}, which
our standard implementation of the HNC does not describe.
Hence, we can offer only a qualitative analysis of the paired phase.
We focus on the symmetric case, $n_1 = n_2 \equiv n$, where all the excitons should pair up into biexcitons.
The biexciton is a bosonic quasiparticle with the effective mass $2m$
and a certain binding energy $E_b$ to be discussed below.
For simplicity, we suppose that no other bound exciton states exist.
At energies much smaller than $E_b$,
the system can be modeled as a single-component Bose liquid with the repulsive interaction potential
\begin{equation}
u_b(r) = u_1(r) + u_2(r) + 2 u_{12}(r)\,.
\label{eqn:u}
\end{equation}
Therefore, the lowest energy excitation mode $E_1(k)$ is again acoustic.
As $k$ increases, the dispersion of $E_1(k)$ should change from linear to the parabolic law
for mass-$2m$ particles: $E_1(k) \simeq \hbar^2 k^2 / 4 m = \epsilon(k) / 2$.
We expect this excitation branch to gradually loose spectral weight as $k$ increases,
as showed schematically by diminishing shading in Fig.~\ref{fig:dilute_spec}(b).
The mode should become essentially extinct at $E_1(k) > E_b$.
The spectral weight gets transferred from this mode to the aforementioned excitation continuum.
The boundary of the continuum
\begin{equation}
	E(k) = E_b + \frac{\hbar^2 k^2}{4 m}
\end{equation}
can be viewed as a Higgs-like mode predicted to exist in superconductors~\cite{Littlewood1982acm} and electron-hole bilayers~\cite{Xue2020hlm}.
The gap in the spectrum is the hallmark of the paired phase.

We used two methods to compute the important energy scale $E_b$.
First, we treated the biexciton as a bound state of a pair of rigid excitons confined to two separate planes.
This problem amounts to solving for the ground state of a single particle of reduced mass $\mu = m / 2$
in a potential $u_{12}(r)$  where $r$ is the distance between the excitons forming the pair.
For the parameters used throughout this article, we found
\begin{equation}
	E_b = 0.286\,\mathrm{meV}.
	\label{eqn:E_b}
\end{equation}
The structure of the biexciton within this approximation is described by wavefunction $\psi_0(r)$
shown in Fig.~\ref{fig:HNCvsPolaron}(d).

The second, more rigorous approach was to solve for the ground state of four particles, two electrons and two holes, in a quadrilayer.
For this task we adopted a stochastic variational method (SVM) previously shown to
be highly accurate for such few-body problems~\cite{Meyertholen2008bit}.
The SVM result $E_b = 0.345\,\mathrm{meV}$ was only $20\%$ larger than that of the simplified first method. [For simplicity, the SVM calculation was done in the limit $c \to 0$ in Eq.~\eqref{eqn:Gaussian}.]

The existence of gapped mode can also be deduced from the BFA.
Indeed, 
let $S_b(k) = 1 + n\, \tilde{h}_b(k)$
be the structure factor and $g_b(r) = 1 + h_b(r)$ be the PCF of ``rigid'' biexcitons, i.e., the single-species bosonic system with the interaction potential $u_b(r)$ [Eq.~\eqref{eqn:u}].
The intraplane structure factors and PCFs of our biexciton superfluid can then be approximated by
\begin{equation}
	S_{\alpha \alpha}(k) = S_b(k)\,,
	\quad
	g_{\alpha \alpha}(r) = g_b(r)\,,
	\quad
	\alpha = 1, 2\,.
	\label{eqn:S_11}
\end{equation}
In turn, the interplane structure factor and PCF are
\begin{align}
	S_{12}(k) &= S_b(k) + \tilde\rho_0(k)\,,
	\label{eqn:S_12}\\
	g_{12}(r) &= g_b(r) + n^{-1} \rho_0(r)\,,
	\label{eqn:g_12}
\end{align}
where $\rho_0(r) = |\psi_0(r)|^2$.
At small $k$, we must have $\tilde{\rho_0}(k) = 1 - b k^2$ with some coefficient $b$.
Equation~\eqref{eqn:E} then implies a finite energy gap $E_{2}(0) = \hbar^2 / (2 m b)$.
The dispersion of the two BFA branches is sketched in
Fig.~\ref{fig:dilute_spec}(b).
The described calculation can be easily performed by the HNC.
We do not show results of such calculations because the BFA
does not accurately predict the true gap $E_2(0) = E_b$ and it does not describe the full distribution of
the spectral weight [shading in Fig.~\ref{fig:dilute_spec}(b)].
As already mentioned, the BFA determines
only the first moments of the eigenmodes of the spectral weight matrix.

In the next Section we examine the case where only one plane is dilute.
We show that such a regime can be studied using methods developed for the polaron problem.

\section{Low density in one of the planes only}
\label{sec:impurity}

\begin{figure}
	\center
	\includegraphics[width=3.0in]{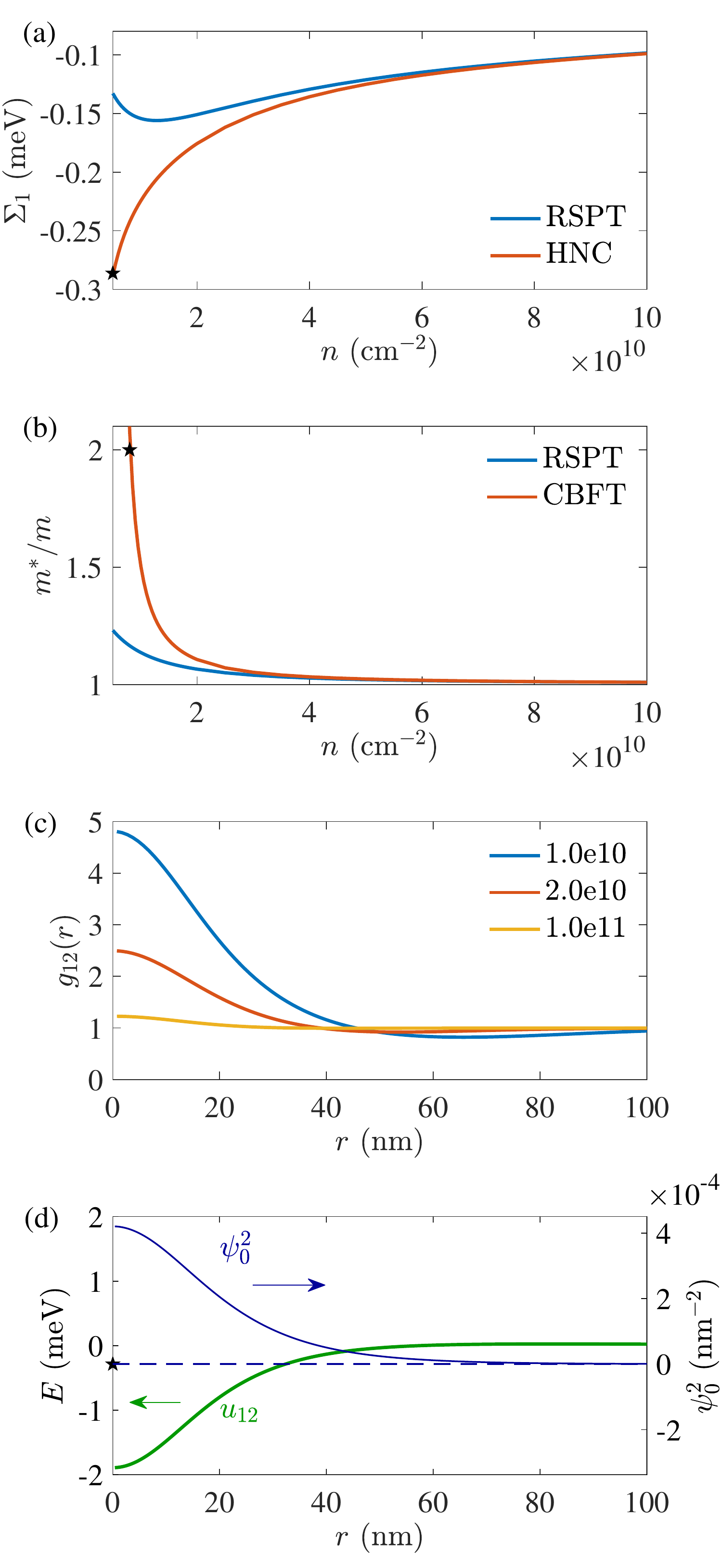}
	\caption{Parameters of a single exciton in plane $1$ interacting with a finite density plane $2$. (a) Chemical potential as a function of exciton density from the RSPT (blue) and HNC (red).
		(b) Effective mass from the RSPT (blue) and CBFT (red).
		(c) Interplane PCFs for densities indicated in the legend (in the units of $\mathrm{cm}^{-2}$).
	(d) The energy [dashed line, Eq.~\eqref{eqn:E_b}] and the wavefunction squared (thin line) of a biexciton bound by the interplane interaction (solid line).}
	\label{fig:HNCvsPolaron}
\end{figure}

The HNC calculations become slowly converging when the exciton densities in both planes is low.
However, if only one plane, say $1$, is dilute, 
so that $n_1 \ll n_2$,
then we can simplify the problem by 
ignoring the interactions in that plane.
The implementation of HNC in such a regime has been considered 
to study impurities in correlated Bose liquids~\cite{Chakraborty1982VTO, owen1981microscopic, pietilainen1983hypernetted}.
Essentially, one needs to take the $n_1 \to 0$ limit of the full HNC equations in Sec.~\ref{sec:HNC}. 
In doing so, it is convenient to redefine $S_{12}$,
\begin{equation}
	S_{12}(k) \equiv  \sqrt{N_2}\, S_{12}^{\mathrm{old}}(k) = n_2 \tilde{h}_{12}(k)
	\label{eqn:S_12new}
\end{equation}
to avoid indeterminate divide-by-zero expressions.
Note that $S_{11}(k) = 1$.
After some algebra, the interplane ``induced potential'' $\tilde{w}_{12}(k)$ in
Eq.~\eqref{eqn:w} reduces to~\cite{Chakraborty1982VTO}
\begin{equation}
	\tilde{w}_{12}(k) = -\frac{(S_{22} - 1) (2 S_{22} + 1)}{2 S_{22}^2}\,
	\frac{S_{12}}{n_2}\, \epsilon(k)
\label{eqn:w_12dilute}
\end{equation}
and the chemical potential $\Sigma_1$ to
\begin{equation}
		\Sigma_1 = -n_2 \tilde{w}_{12}(0) - \frac{1}{2\Omega} \sum_{\vec{k} \neq 0} S_{12}(k) \tilde{w}_{12}(k)\,.
\label{eqn:Sigma_1dilute}
\end{equation}
In the limit of high density, we can evaluate this expression analytically using the formulas
\begin{align}
	S_{22}(k) &\simeq \frac{\epsilon(k)}{\hbar v_2 k}
	\simeq  \left[\frac{\epsilon(k)}{2 n_2 \tilde{u}_{22}(k)}\right]^{1/2} \ll 1\,,
\label{eqn:S_22_high_density}\\
	S_{12}(k) &\simeq -\frac{\tilde{u}_{12}(k)}{\tilde{u}_{22}(k)}\,,
\label{eqn:S_12_high_density}
\end{align}
which follow from Eqs.~\eqref{eqn:E_mf} and \eqref{eqn:E}.
We obtain $\tilde{w}_{12}(0) \simeq -\tilde{u}_{12}(0)$ and
\begin{equation}
	\Sigma_1 \simeq n_2 \tilde{u}_{12}(0) - \frac{1}{2 \Omega} \sum_{\vec{k}}
	\frac{\tilde{u}_{12}^2(k)}{\tilde{u}_{22}(k)}\,,
	\quad n_2 \to \infty.
\label{eqn:Sigma_1_high_density}
\end{equation}
The first term,
which corresponds to a weak interplane repulsion,
can be recognized as the mean-field result.
The second term is due to interplane correlations,
which lead to an effective attraction.
The linear in $n_2$ behavior predicted by Eq.~\eqref{eqn:Sigma_1_high_density}
is in agreement with the numerical evaluation of Eq.~\eqref{eqn:Sigma_1dilute},
which we present in Fig.~\ref{fig:HNCvsPolaron}(a).
Note that $\Sigma_1$ is negative and becomes more negative as $n_2$ decreases, i.e.,
the interplane attraction dominates over repulsion.
Next, the transition to the paired phase can be estimated from the criterion $\Sigma_1 = -E_b$,
marked by the stars in Fig.~\ref{fig:HNCvsPolaron}(a) and (d).
It occurs at $n_2 \approx 0.5 \times 10^{10}\,\mathrm{cm}^{-2}$.
Although these calculations can be extended to still lower $n_2$,
we do not trust such results, and so do not include them in the plot.
The onset of the exciton pairing can also be seen from the PFC shown in Fig.~\ref{fig:HNCvsPolaron}(c).
The maximum of $g_{12}(r)$ at $r = 0$ can be approximated by the biexciton wavefunction squared $\psi_0^2(r)$ [Fig.~\ref{fig:HNCvsPolaron}(d)] multiplied by a coefficient that increases as $n_2$ decreases.
At the lowest density in the plot, this coefficient is close to $n_2^{-1}$,
which is consistent with Eq.~\eqref{eqn:g_12}.

A complementary insight into the problem can be obtained by treating plane $2$ as a harmonic polarizable medium~\cite{hubert2019attractive}.
In this formulation, the excitons of the low-density plane $1$ are analogous to polarons in materials with strong electron-phonon interaction~\cite{mahan2013many}.
Adopting the BFA for the phonon energies $E_2(k)$, we take the effective Hamiltonian of the higher density plane $2$ to be
\begin{align}
	H_2 &= \sum_{\vec{k}} E_2(k) a^{\dagger}_{\vec{k}} a_{\vec{k}}\,,
\label{eqn:H_2}\\
	E_2(k) &= \frac{\epsilon(k)}{S_{22}(k)}\,.
\label{eqn:E_2}
\end{align}
Here $a_{\vec{k}}^\dagger$ and $a_{\vec{k}}$ are the phonon creation and annihilation operators, which are related to the exciton density $\rho_2({\vec{k}})$ in plane $2$ via
\begin{align}
	\rho_2({\vec{k}}) = \sqrt{{N_2} {S_{22}(k)}} \left(a_{\vec{k}} + a^{\dagger}_{-\vec{k}}\right),
	\quad 
	\vec{k} \neq 0\,.
\label{eqn:rho_k}
\end{align}
Note that Eq.~\eqref{eqn:rho_rho} is satisfied.
The effective Hamiltonian for a single exciton with position $\vec{r}$ and
momentum $\vec{p}$ in plane $1$ has the Fr\"ohlich form
\begin{equation}
H_{F} = H_2 + \frac{p^2}{2m} + \frac{1}{\Omega} \sum_{\vec{k} \neq 0} \tilde{u}_{12}(k) e^{i \vec{k}\cdot \vec{r}} \rho_2({\vec{k}}) + n_2 \tilde{u}_{12}(0),
\end{equation}
where the last term represents the $\vec{k} = 0$ contribution.

The two commonly studied properties of a polaron are its energy shift and effective mass $m^*$.
The former is equivalent to our chemical potential $\Sigma_1$ [Eq.~\eqref{eqn:Sigma_l}].
A simple starting point for estimating $\Sigma_1$ and $m^*$ is the Rayleigh-Schr\"odinger perturbation theory (RSPT)~\cite{mahan2013many}.
According to the RSPT, the self-energy $\Sigma(\vec{k})$ of the exciton in plane $1$ is 
\begin{equation}
	\Sigma(\vec{k}) = n_2 \tilde{u}_{12}(0) + \frac{n_2}{\Omega} \sum_{\vec{q} \neq 0}
	\frac{S_{22}(q) u_{12}^2(q)}
	{\epsilon(\vec{k}) - \epsilon(\vec{k} - \vec{q}) - E_2(q)}\,.
	\label{eq:sigmak}
\end{equation}
To compute $\Sigma_1$ and $m^*$ this self-energy is expanded near zero momentum to the order $O(k^2)$:
\begin{align}
	\Sigma(\vec{k}) =  \Sigma_1 - \frac{\hbar^2k^2}{2m}\eta + \ldots\,,
\label{eqn:sigma_expansion}	
\end{align}
which yields
\begin{align}
    \Sigma_1 &= n_2 \tilde{u}_{12}(0) - \frac{n_2}{\Omega} \sum_{\vec{k} \neq 0}
    \frac{\tilde{u}_{12}^2(k) }{\epsilon(k)}
    \frac{S_{22}^2(k)}{1 + S_{22}(k)}\,,
\label{eqn:E_0}\\
    \eta &= \frac{2 n_2}{\Omega} \sum_{\vec{k} \neq 0} \frac{\tilde{u}_{12}^2(k)}{\epsilon^2(k)}\,
     \frac{S_{22}^4(k)}{\left[1 + S_{22}(k)\right]^3}\,.
\label{eqn:eta_RS}
\end{align}
The effective mass is given by
\begin{equation}
	m^* = \frac{m}{1 - \eta}\,.
	\label{eqn:m^*}
\end{equation}
We evaluated these expressions using the structure factor $S_{22}(k)$ supplied by the single-plane HNC and plotted the results in Fig.~\ref{fig:HNCvsPolaron}(a) and (b).
Comparing them with the more reliable HNC calculations,
we can conclude that RSPT can be adequate only at densities above $n_2 \sim 2 \times 10^{10}\,\mathrm{cm}^{-2}$.
In that regime, the mass renormalization is still less than $10\%$, see Fig.~\ref{fig:HNCvsPolaron}(b).
A possible improvement of the RSPT is the Brillouin-Wigner perturbation theory.
Besides $\Sigma_1$ and $m^*$, this theory (Appendix~\ref{sec:BW}) also predicts some intriguing effects such as repulsive polarons.
However, it is difficult to judge how reliable these predictions are.

Another polaron-theory-like approach for calculating the effective mass is the correlated basis functions perturbation theory (CBFT)~\cite{fabrocini2002introduction}.
As described in Appendix~\ref{sec:CBFT}, the CBFT gives
\begin{equation}
    \eta = \frac{1}{2n_2 \Omega} \sum_{\vec{k}} \frac{S_{12}^2(k)}{1 + S_{22}(k)}\,.
    \label{eqn:eta_CBFT}
\end{equation}
The corresponding $m^*$ as a function of $n_2$ is shown in Fig.~\ref{fig:HNCvsPolaron}(b).
One can see that the CBFT and RSPT agree at high density.
[This can also be verified using
Eqs.~\eqref{eqn:S_22_high_density} and \eqref{eqn:S_12_high_density}.]
As $n_2$ decreases, the two perturbation theories diverge from one another.
The CBFT predicts a steep increase of the effective mass $m^*$ of the exciton-polaron at low $n_2$,
see Fig.~\ref{fig:HNCvsPolaron}(b).
From this plot we can infer the paired-phase boundary
using the criterion $m^* = 2 m$.
The corresponding density $n_2\approx 0.8 \times 10^{10}\,\mathrm{cm}^{-2}$ [marked by the star in Fig.~\ref{fig:HNCvsPolaron}(b)] is
somewhat larger than our previous estimate based on $\Sigma_1$ [the star in Fig.~\ref{fig:HNCvsPolaron}(a)].
We speculate that the true phase boundary may be located somewhere in between these two estimates.

\section{Discussion}
\label{sec:conclusions}

Our work was motivated by recent experiments
with e-h-e-h quadrilayer systems~\cite{hubert2019attractive, choksy2021attractive}
that showed evidence for attraction of interlayer excitons.
We modeled the quadrilayer as
a two-plane system of excitons with competing (attractive and repulsive) dipolar interactions.
Using two-species HNC formalism, 
we calculated several zero-temperature properties of the system, 
including the exciton chemical potentials,
in the strongly correlated low-density regime.
Our calculations predict a much weaker attraction effect compared to
what was observed experimentally.
Within our model, the red shift of the chemical potentials and thus the exciton photoluminescence energy
cannot exceed the interplane biexciton binding energy $\sim 0.3\,\mathrm{meV}$ [Eq.~\eqref{eqn:E_b}].
On the other hand, photoluminescence red shifts as large as several $\mathrm{meV}$ have been observed in the experiments~\cite{hubert2019attractive, choksy2021attractive}.
Disorder effects may play some role in explaining this discrepancy. As suggested in Ref.~\onlinecite{choksy2021attractive}, trapping by defects effectively enhances the exciton mass, which in turn increases the polaron energy shift and the biexciton binding. However, such an increase does not seem to be large enough to fully account for the discrepancy between the theory and experiment,
and so further study of this problem is needed.

Another possible line of future investigation of exciton interactions and correlations is probing their 
collective excitations.
To this end we computed the velocities
of the two gapless sound modes that should exist in our system.
We also discussed qualitatively how one of the modes should become gapped in the dilute limit due to the formation of interplane biexcitons.
To the best of our knowledge, collective modes of exciton condensates have not yet been
discovered experimentally.
However, dispersing modes have been detected in (intralayer) exciton-polariton systems~\cite{utsunomiya2008observation, Stepanov2019DRO, Ballarini2020DGW}
using position- and angle-resolved optical spectroscopy.
Additionally, discrete collective resonances have been observed in trapped exciton-polariton liquids~\cite{Estrecho2021LEC} using time-resolved imaging.
The latter technique may be well suited for excitons
because of their long lifetimes, relatively slow dynamics, and
one's ability to manipulate or ``shake''
the traps using external electric gates~\cite{high2009indirect, high2009trapping, kuznetsova2010control}.
If excitons are confined in a trap of size $L$,
the lowest resonant frequency should be of the order of $f \sim {v} / {2L}$
where $v$ is the mode velocity.
If we take $L\sim 50\,\mu \mathrm{m}$ and $v\sim 5\times 10^{6}\,\mathrm{cm}/\mathrm{s}$ (per Sec.~\ref{sec:ground-state}), then $f\sim 5\,\mathrm{GHz}$.
The detection of such a resonant mode does not require complicated ultrafast optics.
Given the detailed geometry of the trap, more accurate computations of the resonant frequencies could be done using our results for the sound velocities as an input.

Note that exciton-polaritons have a very small effective mass and interact weakly,
and so they do not typically form strongly correlated liquids.
In fact, the measured collective mode dispersions are only slightly different from the free-particle
ones and the difference can be adequatly described by the mean-field theory [similar to Eqs.~\eqref{eqn:e_mean_field}, \eqref{eqn:E_mf}].
In contrast, interlayer excitons are usually strongly correlated, so that more sophisticated
many-body theory is needed to study them.
We hope that our work will stimulate further experimental and theoretical work on e-h and e-h-e-h exciton systems.

\acknowledgments

We thank G.~Astrakharchik for providing numerical data for Fig.~\ref{fig:HNCvsMC}(a) and L.~V.~Butov for discussions.
This work is supported by the Office of Naval Research under grant ONR-N000014-18-1-2722.

\appendix

\section{Brillouin-Wigner perturbation theory}
\label{sec:BW}

\begin{figure}
    \centering
    \includegraphics[width=3.5in]{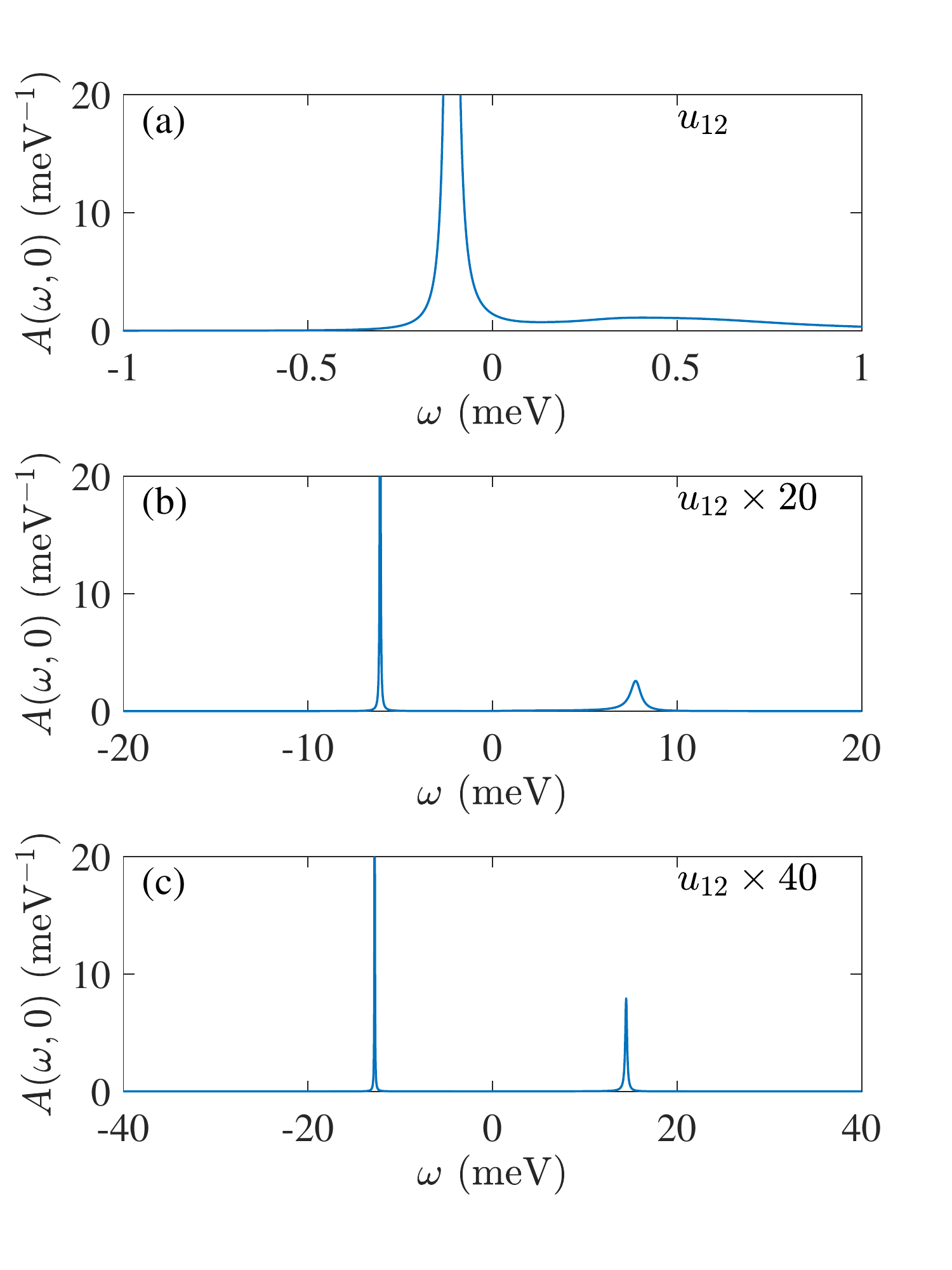}
    \caption{Spectral functions for the particles in the dilute layer for the (a) nominal (b) $10$ times enhanced, and (c) $40$ times enhanced interlayer interaction strength.}
    \label{fig:spectral}
\end{figure}

In this variant of the perturbation theory, the self-energy of the particle in plane $1$ is energy-dependent, i.e., not restricted to be on-shell:
\begin{equation}
    \Sigma(k, E) = \frac{n_2}{\Omega} \sum_{\vec{q}}
    \frac{S_{22}(q) u_{12}^2(q)}{E - \epsilon({\vec{k}-\vec{q}}) - E_2(q) + i\Gamma}\,,
\end{equation}
where $\Gamma > 0$ is a phenomenological damping parameter.
The renormalized dispersion $E(k)$ of this particle is found from the maxima of the spectral function
\begin{equation}
    A(k, E) = -2\,\mathrm{Im} \left[\frac{1}{E - \epsilon(k) - \Sigma(k, E)} \right].
\end{equation}
As an example, we performed the calculation for $n_2 = 0.5\times 10^{10}\,{\mathrm{cm}^{-2}}$, $\vec{k} = 0$, and $\Gamma = 0.05\,\mathrm{meV}$. 
For the nominal interlayer interaction strength, we find a strong peak [Fig.~\ref{fig:spectral}(a)], whose location is near $\Sigma_1$ of the RSPT.
Additionally, at higher energies there is a hint of another weak peak.
The additional dispersion branch associated with this secondary peak is analogous to
``repulsive polarons'' in a system of excitons interacting with a Fermi sea of electrons~\cite{schmidt2012fermi}.
To make the repulsive polaron more apparent, we
repeated the calculation with the interaction potential $u_{12}(r)$ artificially enhanced by a factor of
$20$ and $40$, see Fig.~\ref{fig:spectral}(b) and (c), respectively.

\section{Correlated basis functions perturbation theory}
\label{sec:CBFT}

Let $|0 \rangle$ and $|1 \rangle$ be the ground states of the system containing respectively, zero and one excitons in plane $1$.
The energy difference between these states is the self-energy of the exciton in plane $1$ at zero momentum, $\Sigma_1 = \Sigma(0)$.
The HNC results for $\Sigma_1$ have been discussed in Sec.~\ref{sec:impurity}.
To obtain the exciton effective mass we need to know the self-energy $\Sigma(\vec{k})$ at finite momenta.
Following previous work~\cite{fabrocini2002introduction},
we first consider a set of functions that are direct products of the density wave states in the two planes:
\begin{equation}
	\left|\Psi_{\vec{k}, \vec{q}} \right\rangle = \frac{1}{\sqrt{N_2^{\phantom{1}}}} 
	e^{i (\vec{k} - \vec{q}) \cdot \vec{r}}
	\rho^\dagger_2(\vec{q})| 1\rangle,
\label{eqn:density_basis}
\end{equation}
We refer to the $\vec{q} \neq 0$ functions as the one-phonon states and the $\vec{q} = 0$ function as the zero-phonon state.
Next, we orthogonalize the former with respect to the latter:
\begin{align}
	| \vec{k}, \vec{q}\rangle &= 
	\frac{\left|\Psi_{\vec{k}, \vec{q}}\right\rangle - N_2^{-1} \left|\Psi_{\vec{k}, 0}\right\rangle \left\langle \Psi_{\vec{k}, 0}| \Psi_{\vec{k}, \vec{q}}\right\rangle}
	{\mathcal{N}_{\vec{k}, \vec{q}}^{1 / 2}},
\label{eqn:CBFT_basis}\\
\mathcal{N}_{\vec{k}, \vec{q}} &= \left\langle \Psi_{\vec{k}, \vec{q}}| \Psi_{\vec{k}, \vec{q}}\right\rangle - N_2^{-1} \left|\left\langle \Psi_{\vec{k}, 0}| \Psi_{\vec{k}, \vec{q}}\right\rangle\right|^2,
\end{align}
where $N_2 = \left\langle \Psi_{\vec{k}, 0} |\Psi_{\vec{k}, 0}\right\rangle$ and 
$\mathcal{N}_{\vec{k}, \vec{q}}$ are the normalization factors.
Note that
\begin{align}
	\left\langle\Psi_{\vec{k}, \vec{q}} | \Psi_{\vec{k}, \vec{q}} \right\rangle &= S_{22}(q)\,,
	\\
	\left\langle\Psi_{\vec{k}, 0} | \Psi_{\vec{k}, \vec{q}} \right\rangle &= S_{12}(q)\,.
\end{align}
Ignoring the terms of order $O(N_2^{-1})$,
the matrix elements of the Hamiltonian in this basis are
\begin{align}
	\langle\vec{k}, \vec{q}| H' |\vec{k},\vec{q}\rangle &= \epsilon(\vec{k} - \vec{q})
	 + E_2(q)\,,
\label{eqn:kq_H_kq}	 
\\
	\langle\vec{k}, 0| H' |\vec{k}, \vec{q}\rangle &= -\frac{S_{12}(q)}{\sqrt{N_2 S_{22}(q)}}\,
	\frac{\hbar^2}{2m}\, (\vec{k}\cdot \vec{q}).
\label{eqn:k0_H_kq}	 
\end{align}
The orthogonalization of the basis is important to get Eq.~\eqref{eqn:k0_H_kq}.
The lowest-order perturbative correction to the energy of $\left|\Psi_{\vec{k}, 0} \right\rangle$ state is
\begin{equation}
	\Sigma(\vec{k}) - \Sigma_1
	=  \sum_{\vec{q} \neq 0} \frac{|\langle\vec{k}, 0| H' |\vec{k},\vec{q}\rangle|^2}
	{\epsilon(\vec{k}) - \langle\vec{k},\vec{q}| H' |\vec{k},\vec{q}\rangle }\,.
\label{eqn:Sigma_CBFT}
\end{equation}
Expanding this expression to the order $O(k^2)$, as in
Eq.~\eqref{eqn:sigma_expansion},
we recover Eq.~\eqref{eqn:eta_CBFT} for the mass renormalization parameter $\eta$.
The CBFT can also be done~\cite{fabrocini2002introduction} Brillouin-Wigner style
by replacing $\epsilon(\vec{k})$ with $E$ in Eq.~\eqref{eqn:Sigma_CBFT} but we have not explored that.


\end{document}